# Fringe visibility and distinguishability in two-path interferometer with an asymmetric beam splitter

Yanjun Liu,[1,2] Jing Lu,[1,*] Zhihui Peng,[1] Lan Zhou,[1] and Dongning Zheng[2,3,4]

[1]*Key Laboratory of Low-Dimensional Quantum Structures and Quantum Control of Ministry of Education,
Department of Physics and Synergetic Innovation Center of Quantum Effects
and Applications, Hunan Normal University, Changsha 410081, China*
[2]*Institute of Physics, Chinese Academy of Sciences, Beijing 100190, China*
[3]*CAS Center for Excellence in Topological Quantum Computation and School of Physical Sciences, Beijing 100049, China*
[4]*University of Chinese Academy of Sciences, Beijing 100049, China*

We study the fringe visibility and the distinguishability of a general Mach-Zehnder interferometer with an asymmetric beam splitter. Both the fringe visibility $V$ and the distinguishability $D$ are affected by the input state of the particle characterized by the Bloch vector $\vec{S} = (\vec{S_x}, \vec{S_y}, \vec{S_z})$ and the second asymmetric beam splitter characterized by paramter $\beta$. For the total system is initially in a pure state, it is found that the fringe visibility reaches the upper bound and the distinguishability reaches the lower bound when $\cos\beta = -S_x$. The fringe visibility obtain the maximum only if $S_x = 0$ and $\beta = \pi/2$ when the input particle is initially in a mixed state. The complementary relationship $V^2 + D^2 \leq 1$ is proved in a general Mach-Zehnder interferometer with an asymmetric beam splitter, and the conditions for the equality are also presented.

## I. INTRODUCTION

A single quantum system has mutually exclusive properties, and these characteristics can be converted to each other depending on the method of observation, which is known as Bohr's complementarity principle [1]. The well-known example of the complementarity principle is wave-particle duality. A two-path interferometer, such as Young's double-slit or Mach-Zehnder interferometer (MZI), is used to quantify the wave-particle duality. The wave-like property and the particle-like property are shown by the fringe visibility and the which-path information (WPI) of the interferometer, respectively [2–5]. If the path of the particle is known accurately, the fringe visibility will disappear. The more WPI is obtained, the less the fringe visibility is shown. The complementarity between the fringe visibility and the which-path knowledge has been studied greatly in theory and experiment [3–23].

The complementary relationship between the fringe visibility and the distinguishability has been obtained in the standard MZI [5]. Later, the complementary relationship in a general MZI with an asymmetric beam splitter (BS) has also attracted great attention [24–26]. The asymmetry of the BS1 is equivalent to changing the initial state of the input particles. Since the complementary relationship is established for all the initial state of the input particles, the asymmetry of the BS1 does not affect the complementary relationship. Ref. [25] has proposed that additional *a priori* WPI is introduced when the BS2 is asymmetrical. Unlike the symmetric MZI, the fringe visibility measured at one or the other output port is different in an asymmetric MZI. In this paper, we study the fringe visibility and the distinguishability in a general MZI with an asymmetric BS. It is found that the magnitudes of fringe visibility and the distinguishability are effected by the asymmetric BS and the input state of the particle. The conditions of the upper bound of the fringe visibility and the lower bound of the distinguishability are also obtained, respectively.

The paper is organized as follows. In Sec. II, we introduce the MZI with an asymmetric BS, and study the state evolution of the particle and the the detector in the apparatus. In Sec. III, we obtain the upper bound of the fringe visibility, and present the condition of obtaining the upper bound. In Sec. IV, the lower bound of the distinguishability is found, and the condition for obtaining the lower bound is also proposed. In Sec. V, we made our conclusion.

## II. THE APPARATUS AND THE STATE EVOLUTION

A general MZI consists of two BSs and two phase shifters(PSs) as shown in Fig. 1. The incident particles are split into two paths by the symmetrical beam splitter BS1. Orthogonal normalized states $|a\rangle$ and $|b\rangle$ are used to denote two possible paths, which support a two-dimensional $H_q$. When the particles propagate in these two paths, PS1 and PS2 perform an rotation

$$U_P(\phi) = \exp(-i\phi\sigma_z), \quad (1)$$

on the path qubit, where pauli matrix $\sigma_z = |b\rangle\langle b| - |a\rangle\langle a|$. Finally, these two paths are recombined by the asymmetric beam splitter BS2. The effect of the BS2 on the particle is denoted by

$$U_B(\beta) = \exp(-i\frac{\beta}{2}\sigma_y), \quad (2)$$

which is equivalent to performing a rotation around the $y$ axis by angle $\beta$. The BS2 is symmetrical when $\beta = \pi/2$.

To obtain the WPI, a WPD is placed on path a. When a particle is initially in state

$$\rho_{in}^Q = \frac{1}{2}(1 + S_x\sigma_x + S_y\sigma_y + S_z\sigma_z), \quad (3)$$

*Corresponding author; Electronic address: lujing@hunnu.edu.cn

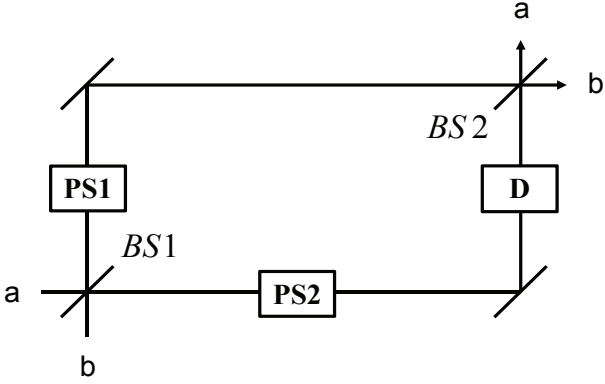

FIG. 1: The schematic sketch of the general Mach-Zehnder interferometer with the second BS asymmetric, a WPD is placed in path a.

go through the MZI, the operator $M = |b\rangle\langle b| \otimes I + |a\rangle\langle a| \otimes U$ is performed on the initial state $\rho_{in}^D$ of the WPD, where $I$ and $U$ are the identical and unitary operator, respectively. In Eq. (3), the Bloch vector $\vec{S} = (\vec{S}_x, \vec{S}_y, \vec{S}_z)$. After the particle has passed through the MZI, the state becomes

$$\begin{aligned}\rho_f &= U_B(\beta) M U_P(\varphi) U_B(\frac{\pi}{2}) \rho_{in}^Q \rho_{in}^D U_B^\dagger(\frac{\pi}{2}) U_P^\dagger(\varphi) M^\dagger U_B^\dagger(\beta) \\ &= \frac{1}{4}(1 - S_x)(1 + \sigma_z \cos\beta + \sigma_x \sin\beta) \otimes \rho_{in}^D \\ &\quad - \frac{1}{4} e^{-i\phi}(S_z - iS_y)(\sigma_z \sin\beta - \sigma_x \cos\beta - i\sigma_y) \otimes \rho_{in}^D U^\dagger \\ &\quad - \frac{1}{4} e^{i\phi}(S_z + iS_y)(\sigma_z \sin\beta - \sigma_x \cos\beta + i\sigma_y) \otimes U\rho_{in}^D \\ &\quad + \frac{1}{4}(1 + S_x)(1 - \sigma_z \cos\beta - \sigma_x \sin\beta) \otimes U\rho_{in}^D U^\dagger. \end{aligned} \quad (4)$$

### III. FRINGE VISIBILITY GAIN VIA GENERAL MZI

The probability that the particle is detected at output port $a$ reads

$$\begin{aligned} p(\phi) &= tr_{QD}[\frac{1}{2}(1 - \sigma_z)\rho_f] \\ &= \frac{1}{2}(1 + S_x \cos\beta) \\ &\quad + \frac{A}{2}\sqrt{\lambda - S_x^2} \sin\beta \cos(\alpha + \gamma + \phi), \end{aligned} \quad (5)$$

where $A = |tr_D(U\rho_{in}^D)|$, $S_x^2 + S_y^2 + S_z^2 = \lambda \leq 1$, $\alpha$ and $\gamma$ are the phases of $S_x + iS_y$ and $tr_D(U\rho_{in}^D)$, respectively. When $\lambda = 1$, the particle is in a pure state, when $\lambda < 1$, the particle is in a mixed state. The fringe visibility, which characterize the wave-like property of the particle, is defined via the probablity in Eq. (5) as

$$V \equiv \frac{max p(\phi) - min p(\phi)}{max p(\phi) + min p(\phi)} = \frac{A \sin\beta}{1 + S_x \cos\beta} \sqrt{\lambda - S_x^2}, \quad (6)$$

where the maximum and minimum is achieved by adjusting $\phi$. We note that the expression of the fringe visibility measured in either output port a or b is different in an asymmetric MZI. Eq. (6) shows that the fringe visibility is effected by the initial state of the particle and the BS2. It is found that for a given $\beta$, more of the waves nature appear when the input particle is in a pure state, i.e. $\lambda = 1$. To demonstrate the effect of the initial state of the particle and the BS2 on fringe visibility, the fringe visibility as a function of the parameter $S_x(\beta)$ for a given $\beta(S_x)$ with $\lambda = 9/25$ or $1$ are shown in Fig. 2. In Fig. 2, we have chosen the parameter $A = 1/3$. In Fig. 2(a) ((c)), we plot the fringe visibility as a function of the parameter $S_x$ for $\lambda = 9/25(1)$ and $\beta = \pi/4, \pi/2, 3\pi/4$ in solid line, dash line and dotted line. From Figs. 2(a) and 2(c) we can obtain that the fringe visibility first increase and then decreases as $S_x$ increase for a given $\beta$. The $S_x$ varies from $-\sqrt{\lambda}$ to $\sqrt{\lambda}$. The value of the fringe visibility is zero when $S_x = \pm\sqrt{\lambda}$, since the particle is determined to propagate only in the $a$ path or the $b$ path, which is corresponding to $S_x = \sqrt{\lambda}$ or $-\sqrt{\lambda}$. The position of the peak changes as $\beta$ changes, and the peak of the fringe visibility appears when $S_x = -\lambda \cos\beta$. For a given $\beta$, the change of $\lambda$ affect the position where the upper bound of the fringe visibility reaches. In the Fig. 2(b) ((d)), we plot the fringe visibility as a function of the parameter $\beta$ for $\lambda = 9/25(1)$ and a given $S_x$. The $S_x = -0.5, 0, 0.5$ are shown by solid line, dash line and dotted line, respectively. The fringe visibility first increase and then decrease as $\beta$ increases for a given $S_x$ are shown in Figs. 2(b) and 2(d). The $\beta$ varies from 0 to $\pi$. The value of the fringe visibility is zero when $\beta = 0$ or $\pi$, since the effect of the BS2 for the particle is full transmission or full reflection, which is corresponding to $\beta = 0$ or $\pi$. The position of the peak changes as $S_x$ changes, and the peak of the fringe visibility appears when $S_x = -\cos\beta$. Figs. 2(b) and 2(d) show that for a given $S_x$, the change of $\lambda$ does not affect the position where the upper bound of the fringe visibility reaches. By comparing Figs. 2(a) and 2(c) or Figs. 2(b) and 2(d), we found that when the input particle is initially in a mixed state, the maximum $\sqrt{\lambda}A$ of the fringe visibility is reached only if $S_x = 0$ and $\beta = \pi/2$, i.e. $S_y^2 + S_z^2 = \lambda$ and the BS2 is symmetrical, and the maximum $A$ of the fringe visibility is reached when the input particle is initially in a pure state with $S_x = -\cos\beta$. It is also found that the upper bound of the fringe visibility changes as $\beta(S_x)$ changes for a given $S_x(\beta)$ when the input particle initially is in a mixed state, and the upper bound of the fringe visibility does not changes as $\beta(S_x)$ changes for a given $S_x(\beta)$ when the input particle initially is in a pure state. Mathematically, by solving the second-order partial derivative of $S_x$ of Eq. (6), it is found that the fringe visibility obtains the upper bound $A\sqrt{\lambda^2 - S_x^2}/\sqrt{\lambda - S_x^2}$ with $S_x = -\lambda \cos\beta$. The fringe visibility reachs the upper bound $A\sqrt{\lambda - S_x^2}/\sqrt{1 - S_x^2}$ with $S_x = -\cos\beta$ by solving the second-order partial derivative of $\beta$ of Eq. (6). By simple calculations, we can obtain the maximum $A$ of the fringe visibility when the input particle initially is in a pure state with $\cos\beta = -S_x$, and the fringe visibility obtain the maximum $\sqrt{\lambda}A$ when the input particle initially is in a mixed state, and the maximum is reached only if $S_x = 0$ and $\beta = \pi/2$.

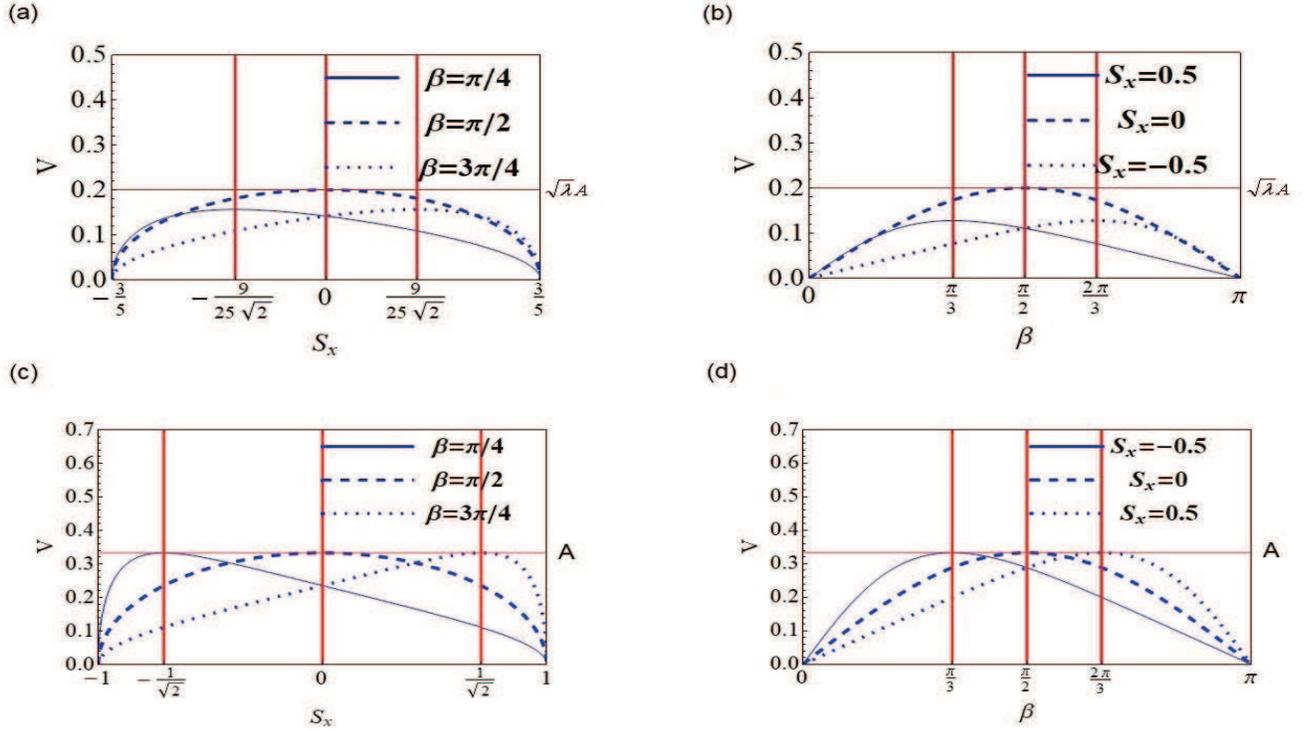

FIG. 2: (Color online). The fringe visibility as a function of $S_x$ for a given $\beta = \pi/4$ (solid line), $\beta = \pi/2$ (dash line), $\beta = 3\pi/4$ (dotted line) with $\lambda = 9/25$ (a) and $\lambda = 1$ (c). The fringe visibility as a function of $\beta$ for a given $S_x = -0.5$ (solid line), $S_x = 0$ (dash line), $S_x = 0.5$ (dotted line) with $\lambda = 9/25$ (b) and $\lambda = 1$ (d). Throughout, we have set $A = 1/3$.

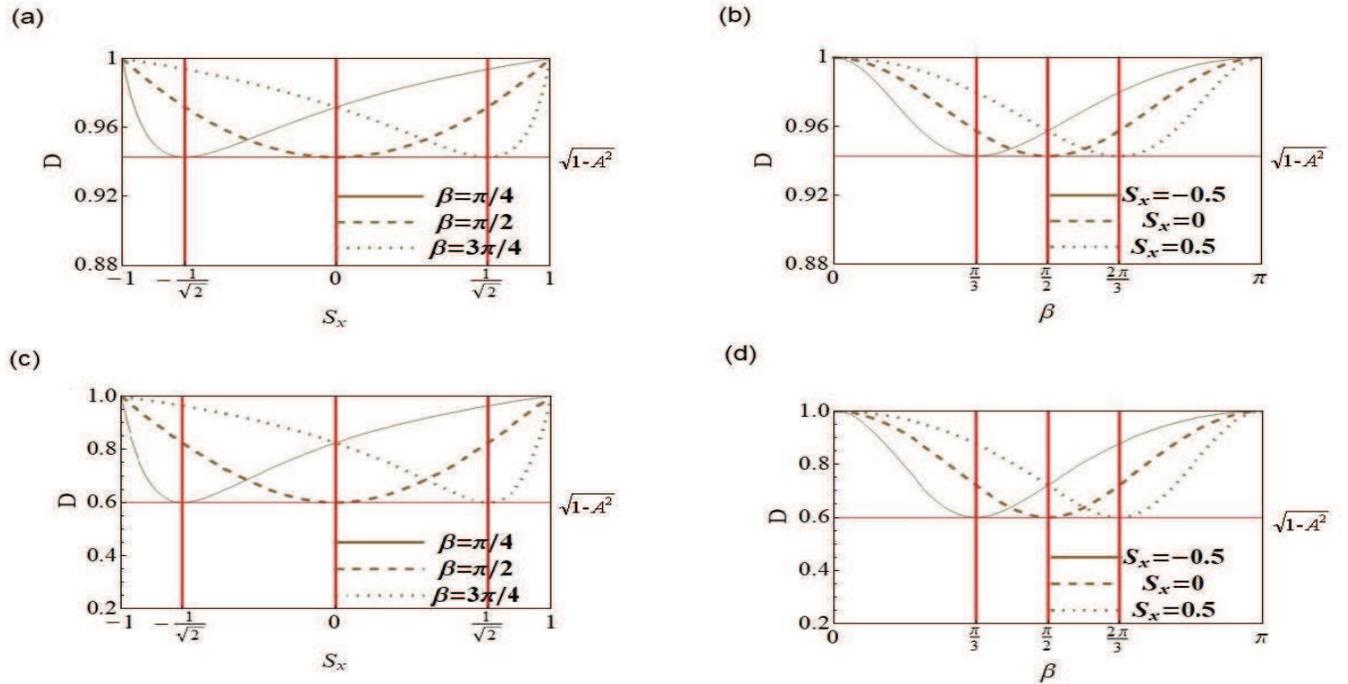

FIG. 3: (Color online). The distinguishability as a function of $S_x$ for a given $\beta = \pi/4$ (solid line), $\beta = \pi/2$ (dash line), $\beta = 3\pi/4$ (dotted line) with $A = 1/3$ (a) and $A = 4/5$ (c). The distinguishability as a function of $\beta$ for a given $S_x = -0.5$ (solid line), $S_x = 0$ (dash line), $S_x = 0.5$ (dotted line) with $A = 1/3$ (b) and $A = 4/5$ (d).

## IV. DISTINGUISHABILITY GAIN VIA GENERAL MZI

To study the particle-like property in the general MZI, the apparatus with four input and output ports has been introduced in Ref. [26]. The state of the WPD was obtained

$$\rho_f^D = \omega_b \rho_{in}^D + \omega_a U \rho_{in}^D U^\dagger, \qquad (7)$$

where

$$\omega_a = \frac{\cos^2\left(\frac{\beta}{2}\right)(1+S_x)}{1+S_x \cos\beta}, \qquad \omega_b = \frac{\sin^2\left(\frac{\beta}{2}\right)(1-S_x)}{1+S_x \cos\beta}. \qquad (8)$$

We note that the probabilities $\omega_a$ snd $\omega_b$ are only dependent on the parameters $S_x$ and $\beta$.

For the convenience of calculation, we assume that the initial state of the WPD is in a pure state denoted by $\rho_{in}^D = |r\rangle\langle r|$. For an arbitrary unitary operator $U$, states $|r\rangle$ and $|s\rangle \equiv U|r\rangle$ are linearly independent, which are either orthogonal state or non-orthogonal state. However, it is well known in quantum measurement theory that non-orthogonal states cannot be accurately distinguished. If conclusive results are made, errors are unavoidable. Thence it is desirable to distinguish these states with minimum probability of error, which is done by the minimum error measurement. Mathematically, the minimum error measurement for two-dimensional Hilbert space is characterized by the projective operators $\Pi_a^D = |M_a\rangle\langle M_a|$ and $\Pi_b^D = |M_b\rangle\langle M_b|$. For state in Eq. (7), the basis of the minimum error measurement read

$$
\begin{aligned}
|M_a\rangle &= -\frac{1}{A_a\sqrt{1-A^2}}\left(\frac{1-\sqrt{1-4\omega_a\omega_b A^2}}{2\omega_a A}\right)|r\rangle + \frac{1}{A_a\sqrt{1-A^2}}U|r\rangle \\
|M_b\rangle &= -\frac{1}{A_b\sqrt{1-A^2}}\left(\frac{1+\sqrt{1-4\omega_a\omega_b A^2}}{2\omega_a A}\right)|r\rangle + \frac{1}{A_b\sqrt{1-A^2}}U|r\rangle,
\end{aligned}
$$
(9)

where

$$
\begin{aligned}
A_a &= \sqrt{\frac{1-4\omega_a\omega_b A^2 - \sqrt{1-4\omega_a\omega_b A^2}(1-2\omega_a A^2)}{2\omega_a^2 A^2(1-A^2)}} \\
A_b &= \sqrt{\frac{1-4\omega_a\omega_b A^2 + \sqrt{1-4\omega_a\omega_b A^2}(1-2\omega_a A^2)}{2\omega_a^2 A^2(1-A^2)}}.
\end{aligned}
$$
(10)

In fact, the basis in Eq. (9) are the eigenstates of the operator $\omega_a U \rho_{in}^D U^\dagger - \omega_b \rho_{in}^D$, which is related to the distinguishability [5]

$$D = Tr_D|\omega_a U \rho_{in}^D U^\dagger - \omega_b \rho_{in}^D| = \sqrt{1-\frac{sin^2\beta(1-S_x^2)A^2}{(1+S_x\cos\beta)^2}}. \quad (11)$$

Here, the magnitude of the distinguishability is determined by parameters $\beta$, $S_x$ and $A$. The fringe visibility is determined by all the components of Bloch vector, but only $S_x$ appears in the expression of the distinguishability, indicating that the distinguishability is not dependent on the $S_y$ and $S_z$. To demonstrate the effect of the initial state of the particle and the BS2 on distinguishability, the distinguishability as a function of the parameter $S_x(\beta)$ for a given $\beta(S_x)$ with $A = 1/3$ or $4/5$ are shown in Fig. 3. In Fig. 3(a) ((c)), we plot the distinguishability as a function of the parameter $S_x$ for $A = 1/3(4/5)$ and $\beta = \pi/4, \pi/2, 3\pi/4$ with solid line, dash line and dotted line. From Figs. 3(a) and 3(c) we can obtain that the distinguishability first decrease and then increase as $S_x$ increases for a given $\beta$. The $S_x$ varies from -1 to 1. The value of the distinguishability is 1 when $S_x = \pm 1$, since the particle is determined to propagate only in the $a$ path or the $b$ path, which is corresponding to $S_x = 1$ or -1. The position of the valley changes as $\beta$ changes, and the valley appears at $S_x = -1/\sqrt{2}$ when $\beta = \pi/4$, $S_x = 0$ when $\beta = \pi/2$ and $S_x = 1/\sqrt{2}$ when $\beta = 3\pi/4$ in Figs. 3(a) and 3(c). Figs. 3(a) and 3(c) show that for a given $\beta$, the change of $A$ does not affect the position where the lower bound of the distinguishability reaches. In the Fig. 3(b) ((d)), we plot the distinguishability as a function of the parameter $\beta$ for $A = 1/3(4/5)$ and a given $S_x$. The $S_x = -0.5, 0, 0.5$ are shown by solid line, dash line and dotted line, respectively. The distinguishability first decrease and then increase as $\beta$ increases for a given $S_x$ are shown in Figs. 3(b) and 3(d). The $\beta$ varies from 0 to $\pi$. The value of the distinguishability is 1 when $\beta = 0$ or $\pi$, since the effect of the BS2 for the particle is full transmission or full reflection, which is corresponding to $\beta = 0$ or $\pi$. The valley appears at $\beta = \pi/3$ when $S_x = -0.5$, $\beta = \pi/2$ when $S_x = 0$ and $\beta = 2\pi/3$ when $S_x = 0.5$ in Figs. 3(b) and 3(d). Figs. 3(b) and 3(d) show that for a given $S_x$, the change of $A$ does not affect the position where the lower bound of the distinguishability reaches. By comparing Figs. 3(a) and 3(c) or Figs. 3(b) and 3(d), we find that the lower bound of the distinguishability is determined by $A$, and the position of the lower bound is determined by $S_x$ and $\beta$. It is also found that the distinguishability obtains the lower bound $A$ when $\cos\beta = -S_x$ in Fig. 3. When $\cos\beta = -S_x$, $\omega_a = \omega_b = 1/2$ in Eq. (7). In this case, the amount of information obtains the upper bound and the distinguishability obtains the lower bound. Mathematically, we can obtain the minimum $\sqrt{1-A^2}$ of the distinguishability when $\cos\beta = -S_x$ by solving the second-order partial derivative of $S_x$ or $\beta$ of Eq. (11).

## V. CONCLUSION

We have investigated the effect of the initial state of the particle and the BS2 on both the fringe visibility and the distinguishability. The fringe visibility characterized by $V$ obtains the upper bound $A = |tr_D(U\rho_{in}^D)|$ when the total system initially in a pure state and $\cos\beta = -S_x$, and the maximum $A$ is determined by the initial state of the detector and the unitary operator performed on it. The upper bound of the fringe visibility is related to $A$, $\lambda$ and $S_x$ when the input particle is initially in a mixed state, and the condition for obtaining the upper bound of the fringe visibility for a given $\beta$ is different from that obtained for a given $S_x$. The fringe visibility obtain the maximum $\sqrt{\lambda}A$ when the input particle is initially in a mixed state, and the maximum is reached only if $S_x = 0$ and $\beta = \pi/2$. The distinguishability is characterized by the

$D$, which is obtained via minimum error discrimination of the state of the detector. The lower bound $\sqrt{1-A^2}$ of the distinguishability can be achieved when $\cos\beta = -S_x$. We find that the lower bound of the distinguishability is determined by $A$, and the position of the lower bound are determined by $S_x$ and $\beta$. By equations (6) and (11), the complementary relationship $V^2 + D^2 = 1 - [A^2 \sin^2\beta(1-\lambda)]/(1+S_x \cos\beta)^2$ when the initially detector is in a pure state and the state of the initially particle is arbitrary. The complementary relationship $V^2 + D^2 = 1$ in the following situations. (1) The effect of the BS2 for the particle is full transmission or full reflection, which is corresponding to $\beta = 0$ or $\pi$. (2) The total system is initially in a pure state, which is corresponding to $\lambda = 1$. (3) The states of the detector $|r\rangle$ and $|s\rangle$ are orthogonal state, which is corresponding to $A = 0$.

**Acknowledgments**

This work was supported by NSFC Grants Nos. 11434011, 11575058 and 61833010; Hunan Province "Science and Technology Innovation Platform and Talent Plan" Excellent Talent Award No. 2017XK2021; Science Funds from the Ministry of Science and Technology of China Grant Nos. 2017YFA0304300 and 2016YFA0300601; Strategic Priority Research Program of the Chinese Academy of Sciences Grant No. XDB28000000.